\documentclass[twocolumn,showpacs,amsmath,amssymb,aps,pra,floatfix,10pt]{revtex4-1}

\usepackage[latin1]{inputenc}
\usepackage[american]{babel}
\usepackage[T1]{fontenc}
\usepackage{amsmath}
\usepackage{upgreek}
\usepackage{mhchem}
\usepackage{graphicx}
\usepackage{multirow}
\usepackage{physics}
\usepackage{changes}
\usepackage{xcolor}

\begin{document}

\title{Skyrme-type nuclear interaction as a tool for calculating the finite nuclear size correction to atomic energy levels and the bound-electron $\boldsymbol{g}$ factor}

\date{\today}

\author{Igor~A.~Valuev}
\email[Email: ]{igor.valuev@mpi-hd.mpg.de} 
\affiliation{Max-Planck-Institut f\"{u}r Kernphysik, Saupfercheckweg 1, 69117 Heidelberg, Germany}
\author{Zolt\'{a}n~Harman}
\affiliation{Max-Planck-Institut f\"{u}r Kernphysik, Saupfercheckweg 1, 69117 Heidelberg, Germany}
\author{Christoph~H.~Keitel}
\affiliation{Max-Planck-Institut f\"{u}r Kernphysik, Saupfercheckweg 1, 69117 Heidelberg, Germany}
\author{Natalia~S.~Oreshkina}
\email[Email: ]{natalia.oreshkina@mpi-hd.mpg.de} 
\affiliation{Max-Planck-Institut f\"{u}r Kernphysik, Saupfercheckweg 1, 69117 Heidelberg, Germany}

\begin{abstract}
A state-of-the-art approach for calculating the finite nuclear size correction to atomic energy levels and the bound-electron $g$~factor is introduced and demonstrated for a series of highly charged hydrogen-like ions.
Firstly, self-consistent mean-field calculations based on the Skyrme-type nuclear interaction are employed in order to produce a realistic nuclear proton distribution.
In the second step, the obtained nuclear charge density is used to construct the potential of an extended nucleus, and the Dirac equation is solved numerically.
The ambiguity in the choice of a Skyrme parametrization is supressed by fine-tuning of only one parameter of the Skyrme force in order to accurately reproduce the experimental values of nuclear radii in each particular case.
The homogeneously charged sphere approximation, the two-parameter Fermi distribution and experimental nuclear charge distributions are used for comparison with our approach, and the uncertainties of the presented calculations are estimated.
In addition, suppression of the finite nuclear size effect for the specific differences of $g$ factors is demonstrated.   
\end{abstract}

\maketitle

\section{Introduction}\label{sec:intro}
Highly charged ions represent one of the simplest and most well-understood physical systems, and yet they still continue to provide an extremely rich scope of opportunities for fundamental research. 
They have been extensively used in past years for various high-precision tests of quantum electrodynamics, making it one of the most well-tested theories in physics \cite{Draganic2003, Gumberidze2005, Shabaev2006109, Volotka_PRL_2012, Shabaev_JPhysCh_2015}.
Such a high accuracy has also been employed for a precise determination of the electron mass \cite{Sturm_nature_2014} and it has been proposed to be used for determination of the fine-structure constant \cite{PhysRevLett.96.253002, Volotka_PRL_2014, PhysRevLett.116.100801} and even for search of its hypothetical variation \cite{Andreev_PRL_2005, Berengut2010, Windberger_PRL_2015,  Oreshkina_PRAR_2017, bekker2019detection}.
Moreover, comparison between the experimental and theoretical results can be used to test theories beyond the Standard Model by setting bounds on parameters of new hypothetical forces~\cite{debierre2019g}.
Among various achievements in this field, the most prominent ones include measurements and calculations of the bound-electron $g$~factor in highly charged ions to an extraordinary level of precision \cite{carbon_gfactor, oxygen_gfactor, silicon_gfactor, argon_gfactor, silicon_li_gfactor, Cakir2020}.
For all types of high-presicion spectroscopic measurements of highly charged ions \cite{Beiersdorfer1998, TokyoEBIT, Egl2019} the essential and most fundamental quantities are atomic energy levels and corresponding transition energies which are needed to be known to a high level of accuracy from the theoretical side.

As the experimental precision is being improved, nuclear structure effects are also becoming observable and thus have to be calculated with an increasing accuracy.
The largest correction of this kind is due to the finite nuclear size (FNS) effect.
Analytical expressions for the FNS effect were presented in \cite{Shabaev_1993, GLAZOV2002408, Karshenboim2005, Karshenboim2018}.
The FNS correction can also be calculated with a higher accuracy numerically by using the Fermi distribution as a model for nuclear charge density \cite{BEIER200079}. 
However, even this model does not describe any fine details of nuclear charge distributions that are unique for each nucleus.
Hence, in order to perform more precise calculations of the FNS correction, it is necessary to use a more realistic nuclear-structure description and go beyond the simple Fermi model.
As for other nuclear-structure corrections, we note that significant improvements in the evaluation of nuclear deformation and nuclear polarization effects have been made in recent years \cite{NEFIODOV1996227, Kozhedub2008, nucl_shape2012, nucl_pol_volotka, nucl_shape2019}.

In this paper we present calculations of the FNS correction to atomic energy levels and the bound-electron $g$~factor based on a more detailed description of nuclear charge distributions. 
The nuclear charge densities are calculated in the framework of the Hartree-Fock method based on the Skyrme-type nuclear interaction with adjustable parameters.
We employ the \verb|skyrme_rpa| program for this purpose \cite{COLO2013142}. 
The obtained data is then used to construct the potential of an extended nucleus and numerically solve the Dirac equation for an electron bound in this potential.
The theoretically calculated nuclear charge densities are in a good agreement with the experimental ones. These results pave the way for a more accurate description of nuclear-structure effects in atomic systems.

The paper is organized as follows. 
After a brief description of the computational method we discuss the numerical results and their dependence on the Skyrme parameters.
We compare our results to the FNS corrections obtained  by using experimental nuclear charge distributions as well as simpler charge density models, such as the homogeneously charged sphere approximation and the Fermi distribution. 
Then we estimate the uncertainties of our calculations and also demonstrate suppression of the FNS effect for the specific differences \cite{spec_dif2002,PhysRevLett.96.253002} of $g$~factors.

Relativistic system of units ($\hbar = c = 1$) and Heaviside charge units ($\alpha = e^2/4\pi, e<0$) are used throughout the paper.
Three-vectors are denoted by bold letters.  

\section{Computational method}\label{sec:method}

\subsection{Skyrme interaction and nuclear charge density}\label{sec:density}
In its standard form, the Skyrme interaction between two nucleons with spatial coordinates ${\bf r}_1$ and ${\bf r}_2$ can be expressed as \cite{CHABANAT1998231}: 
\begin{align}
V({\bf r}_1, {\bf r}_2) & = t_0 \left( 1 + \chi_0 P_\upsigma \right) \delta({\bf r}) \notag \\ 
& + \dfrac{1}{2} t_1 \left( 1 + \chi_1 P_\upsigma \right) \left[{\bf P}^{\dag2} \delta({\bf r}) + \delta({\bf r}) {\bf P}^2 \right] \notag \\
& + t_2 \left( 1 + \chi_2 P_\upsigma \right) {\bf P}^\dag \cdot \delta({\bf r}) {\bf P} \notag \\
& + \dfrac{1}{6} t_3 \left( 1 + \chi_3 P_\upsigma \right) \rho^\lambda ({\bf R}) \delta({\bf r}) \notag \\
& + iW_0 \left( \boldsymbol{\upsigma}_1 + \boldsymbol{\upsigma}_2 \right) \cdot \left[{\bf P}^\dag \times \delta({\bf r}) {\bf P} \right],
\end{align}
where ${\bf r} = {\bf r}_1 - {\bf r}_2$, 
${\bf R} = \dfrac{1}{2} \left( {\bf r}_1 + {\bf r}_2 \right)$,
${\bf P} = \dfrac{1}{2i} \left( \nabla_1 - \nabla_2 \right)$ 
(${\bf P}^\dag$ acts to the left),
$P_\upsigma = \dfrac{1}{2} \left( 1 + \boldsymbol{\upsigma}_1 \cdot \boldsymbol{\upsigma}_2 \right)$, 
$\upsigma_i$ with $i \in \{ 1,2,3 \}$ are the Pauli spin matrices, and $\rho$ is the total nucleon density.
Here we note that $t_j$, $\chi_j$ ($j~\in~\{ 0,1,2,3 \}$), $W_0$ and $\lambda$ are adjustable parameters of the Skyrme force~\cite{COLO2013142}.

Next, in order to derive the Hartree-Fock (HF) equations, single-particle wave functions $\{ \phi^q_i(x) \}$ are introduced, where $x$ denotes the set of spatial and spin coordinates, and the superscript $q$ is used to distinguish between the neutron ($q=``n$'') and proton ($q=``p$'') orbitals.
The many-body ground-state wave function is built out of these functions as a Slater determinant, and then by means of the variational principle one can obtain the HF equations of the general form: 
\begin{equation}
\widehat{H}(x,\phi^q_i(x)) \phi^q_i(x) = \epsilon_i \phi^q_i(x),
\end{equation}
where the Hamiltonian $\widehat{H}$ itself depends on the single particle wave functions.
The explicit form of the equations as well as their detailed derivation can be found in various articles, for example in \cite{Vautherin_PhysRevC.5.626}. 
The HF equations are solved iteratively until self-consistency to a predefined accuracy is achieved.

The obtained orbitals can then be used to construct point nucleon densities, in particular, the proton density:
\begin{equation}
\rho_p({\bf r}) = \sum_{i,\upsigma} |\phi^p_i({\bf r},\upsigma)|^2.
\end{equation}
Finally, in order to obtain the nuclear charge distribution, the proton density is convoluted with the Gaussian form factor $f_p(r)$ to allow for the finite size of the proton \cite{Vautherin_PhysRevC.5.626}:
\begin{align}
f_p(r) & = \dfrac{1}{\left( r_0 \sqrt{\pi} \right)^3} e^{-r^2/r^2_0} \ , \ r_0=0.65 \ \mathrm{fm}, \\
& \rho_c({\bf r}) = \int f_p({\bf r}-{\bf r'}) \rho_p({\bf r'}) \, \mathrm{d}^3{\bf r'}.
\end{align}
We note here that in the following we assume spherical symmetry of nuclear charge distributions. 

Other expressions for the nuclear charge density $\rho_c(r)$ are used in this paper for comparison purposes, and they include:
\vskip 5pt
\noindent 1) the homogeniously charged sphere approximation (``Sphere''):
\begin{equation}
\label{eq:sphere}
\rho_c(r) = 
   \begin{cases}
   \rho_0^{\mathrm{sphere}} & \mathrm{for} \ 0 \leq r \leq \sqrt{\dfrac{5}{3} \langle r^2 \rangle}, \\
   0 & \mathrm{otherwise};
   \end{cases}
\end{equation}

\noindent 2) Fermi distribution (``Fermi''):
\begin{equation}
\label{eq:Fermi}
\rho_c(r) = \dfrac{\rho_0^{\mathrm{Fermi}}}{1+e^{(r-c)/a}},
\end{equation}
with the radius parameter $c$ and the diffuseness parameter $a = (2.3 / 4 \, \mathrm{ln}3) \; \mathrm{fm} $ \cite{BEIER200079};

\noindent 3) model-independent analyses of experimental scattering data \cite{DEVRIES1987495}: 

\noindent a) expansion into a sum of spherical Bessel functions $j_{0}$ of order zero (``Bessel''):
\begin{equation}
\label{eq:Bessel}
\rho_c(r) = 
   \begin{cases}
   \sum_{\nu} \limits a_{\nu} j_0 \left( \nu \pi r/R \right) & \mathrm{for} \ 0 < r \leq R, \\
   0 & \mathrm{otherwise},
   \end{cases}
\end{equation}
where $R$ is the cutoff radius;

\noindent b) expansion into a sum of Gaussians (``Gauss''):
\begin{align}
\label{eq:Gauss}
& \rho_c(r) = \sum_i A_i \left( e^{- \left[ (r-R_i)/ \gamma \right]^2}
                              + e^{- \left[ (r+R_i)/ \gamma \right]^2}  \right), \\
A_i & = Q_i \left[ 2 \pi^{3/2} \gamma^3 \left( 1 + 2R^2_i / \gamma^2 \right) \right]^{-1}, \quad \sum_i Q_i = 1, \notag                            
\end{align}
where $R_{i}$ and $Q_{i}$ are the positions and the amplitudes of the Gaussians, respectively, and the parameter $\gamma$ is related to the root-mean-square radius $R_{\mathrm{G}}$ of the Gaussians as follows: $R_{\mathrm{G}} = \gamma \sqrt{3/2}$.

In this paper, the constants $\rho_0^{\mathrm{sphere}}$ and $\rho_0^{\mathrm{Fermi}}$ as well as the coefficients $a_{\nu}$ and $Q_i$ in Eqs.~\eqref{eq:sphere}~--~\eqref{eq:Gauss} are chosen to fulfil the following normalization condition: $4\pi \int_{0}^{\infty} \rho_c(r) r^{2} \, \mathrm{d}r = 1$.

\subsection{Dirac equation}\label{sec:Dirac}
Once the nuclear charge density $\rho_c(r)$ is known, one can construct the potential describing the interaction between an electron and the nucleus as follows \cite{reiher2009relativistic}:
\begin{align}
V(r) = \dfrac{-4\pi \alpha Z}{r} \int_{0}^{r} & \rho_c(r') r'^{2} \, \mathrm{d}r' \notag \\
& - 4\pi \alpha Z \int_{r}^{\infty} \rho_c(r')r' \, \mathrm{d}r',
\end{align}
where $Z$ is the nuclear charge, and $\alpha$ is the fine structure constant.
This potential then enters the Dirac equation which determines the energy levels $E$ and the four-component wave functions $\psi ({\bf r})$ of a bound electron \cite{greiner2000relativistic}:
\begin{equation}
\label{eq:Dirac}
\left[ \boldsymbol{\alpha} \cdot {\bf p} + \beta m_{e} + V(r) \right] \psi ({\bf r}) = E \psi ({\bf r}),
\end{equation}
where $\boldsymbol{\alpha}$ and $\beta$ are the usual Dirac matrices, and $m_{e}$ is the electron mass.

For an arbitraty central potential the electron wave function splits into radial and angular parts as:
\begin{equation}
\label{eq:Psi}
\psi_{n \kappa m}({\bf r}) = \dfrac{1}{r} \left( 
   \begin{array}{c}
   G_{n \kappa}(r)\Omega_{\kappa m}(\theta,\varphi) \\[3pt]
   iF_{n \kappa}(r)\Omega_{-\kappa m}(\theta,\varphi)
   \end{array}
\right),   
\end{equation}
where $n$ is the principal quantum number, $\kappa$ is the relativistic angular momentum quantum number, and $m$ is the total magnetic quantum number. The spherical spinors $\Omega_{\pm \kappa m}(\theta,\varphi)$ are the same for any central potential and are well known \cite{johnson2007atomic}. 
Hence, the problem can be reduced to the following set of radial Dirac equations:
\begin{align}
\dfrac{\mathrm{d}G}{\mathrm{d}r} + \dfrac{\kappa}{r}G(r) - \left[ m_{e}-V(r) \right] F(r) & = EF(r), \notag \\
-\dfrac{\mathrm{d}F}{\mathrm{d}r} + \dfrac{\kappa}{r}F(r) + \left[ m_{e}+V(r) \right] G(r) & = EG(r),
\end{align}
where the radial functions $G(r)$ and $F(r)$ satisfy the normalization condition: $\int_{0}^{\infty} \left[ G(r)^{2} + F(r)^{2} \right] \, \mathrm{d}r = 1$.

The radial wave functions $G(r)$ and $F(r)$ can then be found analytically for the Coulomb potential \cite{greiner2000relativistic} or in general case numerically, for example, by expanding them in terms of B-splines and solving the resulting generalized matrix eigenvalue equations \cite{johnson2007atomic}. In our basis-set numerical calculations of the radial wave functions we used the dual-kinetic-balance approach \cite{DKB}. 

In order to obtain the FNS correction to atomic energy levels, the numerically calculated values $E_{\rm{ext}}[n\kappa]$ (in the case of an extended nucleus) are compared to the exact analytical solution $E_{\mathrm{point}}[n\kappa]$ for the Coulomb potential $V(r) = -Z \alpha/r$ (i.e. point-like nucleus):
\begin{align}
& \Delta E_{\mathrm{FNS}}[n\kappa] = E_{\rm{ext}}[n\kappa] - E_{\mathrm{point}}[n\kappa], \\
E_{\mathrm{point}}[n\kappa] & = m_{e} \left[ 1 + \dfrac{(Z\alpha)^2}{\left( n-|\kappa|+\sqrt{\kappa^2-(Z\alpha)^2} \right)^2} \right]^{-1/2}. \notag
\end{align}

\subsection{Bound-electron $\boldsymbol{g}$ factor}\label{sec:gfactor}
Most generally, a $g$~factor relates the electron's magnetic moment $\boldsymbol{\mu}$ (in units of Bohr magneton $\mu_{\rm{B}}=|e|/2m_{e}$) to its angular momentum $\boldsymbol{M}$:
\begin{equation}
\dfrac{\boldsymbol{\mu}}{\mu_{\rm{B}}} = -g \boldsymbol{M}, \ \mathrm{e.g.} \ \
\dfrac{\boldsymbol{\mu_{l}}}{\mu_{\rm{B}}} = -g_{l} \boldsymbol{l} \ \ \mathrm{and} \ \ \dfrac{\boldsymbol{\mu_{s}}}{\mu_{\rm{B}}} = -g_{s} \boldsymbol{s},
\end{equation}
where $\boldsymbol{l}$ is the orbital angular momentum, and $\boldsymbol{s}$ is the spin angular momentum. In the Dirac theory, i.e. without taking into account the radiative corrections, $g_{s}=2$ for a free electron, and $g_{l}$ is known to be exactly 1 \cite{greiner2000quantum}.

Thus, the interaction Hamiltonian $\widehat{H}_{\mathrm{int}}$ for an electron in an external homogeneous magnetic field $\boldsymbol{B}=(0,0,B_{z})$ can be expressed as:
\begin{equation}
\widehat{H}_{\mathrm{int}} = -\boldsymbol{\mu}_{\mathrm{total}} \boldsymbol{\cdot} \boldsymbol{B} = \mu_{\mathrm{B}}(g_{l} \boldsymbol{l} + g_{s} \boldsymbol{s}) \boldsymbol{\cdot} \boldsymbol{B}.
\end{equation} 
The corresponding first-order Zeeman splitting $\Delta E$ can then be written by introducing a new $g$~factor, also called Land\'{e} $g$~factor: 
\begin{equation}
\label{eq:Zeeman1}
\Delta E = \expval{\widehat{H}_{\mathrm{int}}}{n \kappa m} = g \mu_{\rm{B}} B_{z} m,
\end{equation}
We note that Eq. \eqref{eq:Zeeman1} is written in such a way to have the same form as for the simpler case where $l=0$, and it can be considered as a definition of the Land\'{e} $g$~factor of a bound electron.

On the other hand, the electromagnetic four-potential $A^{\mu}$ can be chosen in the form $( 0, \, \boldsymbol{A}(\boldsymbol{r})=[ \boldsymbol{B} \times \boldsymbol{r} ]/2 )$, and an application of the minimal coupling principle to the Dirac equation~\eqref{eq:Dirac} implies:
\begin{equation}
\widehat{H}_{\mathrm{int}}' = -e \boldsymbol{\alpha} \boldsymbol{\cdot} \boldsymbol{A}(\boldsymbol{r}) = |e| \boldsymbol{\alpha} \boldsymbol{\cdot} \boldsymbol{A}(\boldsymbol{r}).
\end{equation}
In this way, first-order perturbation theory gives:
\begin{align}
\label{eq:Zeeman2}
\Delta E & = \dfrac{|e|}{2} \expval{\boldsymbol{\alpha} \boldsymbol{\cdot} [ \boldsymbol{B} \times \boldsymbol{r} ]}{n \kappa m} \notag \\
& = \dfrac{|e|}{2} B_{z} \expval{[ \boldsymbol{r} \times \boldsymbol{\alpha} ]_{z}}{n \kappa m}.
\end{align} 

A calculation of the matrix element in Eq. \eqref{eq:Zeeman2} using the wave functions of the form \eqref{eq:Psi} \cite{rose1961relativistic} and then taking into account Eq. \eqref{eq:Zeeman1} yields the following general formula for the $g$~factor:
\begin{equation}
\label{eqn:gext}
g_{\mathrm{ext}} [n \kappa] = \dfrac{2\kappa m_{e}}{j(j+1)} \int_{0}^{\infty} G_{n \kappa}(r)F_{n \kappa}(r)r \, \mathrm{d}r,
\end{equation}
where $j=|\kappa|-1/2$ is the total angular momentum quantum number.
 
In the case of the Coulomb potential $V(r) = -Z \alpha/r$ an~exact analytical calculation can be performed \cite{zapryagaev1979zeeman}, and the result reads:
\begin{equation}
\label{eqn:gpoint}
g_{\mathrm{point}} [n \kappa] = \dfrac{\kappa}{j(j+1)} \left( \kappa \dfrac{E_{\mathrm{point}} [n\kappa]}{m_{e}} - \dfrac{1}{2} \right).
\end{equation}

Finally, the FNS correction to the $g$ factor for a state $n \kappa$ is obtained by taking the difference between (\ref{eqn:gext}) and (\ref{eqn:gpoint}):
\begin{equation}
\Delta g_{\mathrm{FNS}} [n \kappa] = g_{\mathrm{ext}} [n \kappa] - g_{\mathrm{point}} [n \kappa].
\end{equation}

Other contributions to the $g$~factor are summarized e.g. in Ref. \cite{BEIER200079, ShabaevReview2015, Harman_2018}.

\section{Results and discussion}\label{sec:results}

\subsection{Choice of Skyrme parametrization}\label{sec:parameters}

\begin{table*}[t]
\caption{Comparison between the parameters $t_1$, $\chi_0$ and $\chi_3$ from the LNS, SLy5 and SKP Skyrme parameter sets as~well~as the corresponding calculated values of RMS nuclear radius of \ce{^{208}_{82}Pb} nucleus. The FNS corrections to the ground-state energy $\Delta E_{\mathrm{FNS}}[1s_{1/2}]$ (in units of electron's rest energy) and $g$~factor $\Delta g_{\mathrm{FNS}}[1s_{1/2}]$ for hydrogen-like lead~\ce{^{208}_{82}Pb^{81+}} are presented in the last two columns. For comparison, the results for the homogeneously charged sphere approximation are also included in the last row.}
\label{table:parameters}
{\renewcommand{\arraystretch}{1.5}
\renewcommand{\tabcolsep}{0.3cm}
\begin{tabular}{ccccccc}
\hline \hline
Parameter set & $t_1$ & $\chi_0$ & $\chi_3$ & $\sqrt{\langle r^2 \rangle}$, fm & $\Delta E_{\mathrm{FNS}}[1s_{1/2}] \times 10^4$ & $\Delta g_{\mathrm{FNS}}[1s_{1/2}] \times 10^4$ \\
\hline
LNS & 266.735 & 0.06277 & -0.03413 & 5.3238 & 1.2483 & 4.3014 \\
SLy5 & 483.13 & 0.778 & 1.267 & 5.5072 & 1.3169 & 4.5369 \\
SKP & 320.62 & 0.29215 & 0.18103 & 5.5242 & 1.3234 & 4.5590 \\
\hline
Sphere & - & - & - & 5.5012 & 1.3172 & 4.5380 \\
\hline \hline
\end{tabular}}
\end{table*}

\begin{table*}[!hbtp]
\caption{Modifications of the $t_0$ Skyrme parameter within the LNS, SLy5 and SKP parametrizations and the corresponding FNS corrections to the ground-state energy $\Delta E_{\mathrm{FNS}}[1s_{1/2}]$ (in units of electron's rest energy) and $g$~factor $\Delta g_{\mathrm{FNS}}[1s_{1/2}]$ for hydrogen-like lead~\ce{^{208}_{82}Pb^{81+}}.}
\label{table:adjust}
{\renewcommand{\arraystretch}{1.5}
\renewcommand{\tabcolsep}{0.3cm}
\begin{tabular}{cr@{}lcc}
\hline \hline
Parameter set & \multicolumn{2}{c}{Change in $t_0$} & $\Delta E_{\mathrm{FNS}}[1s_{1/2}] \times 10^4$ & $\Delta g_{\mathrm{FNS}}[1s_{1/2}] \times 10^4$ \\
\hline
LNS & -2484.&97 $\rightarrow$ -2454.60 (1.22\%) & 1.3148 & 4.5296 \\
SLy5 & -2484.&88 $\rightarrow$ -2486.12 (0.05\%) & 1.3147 & 4.5291 \\
SKP & -2931.&70 $\rightarrow$ -2935.95 (0.15\%) & 1.3147 & 4.5291 \\
\hline \hline
\end{tabular}}
\end{table*}

First, let us discuss the influence of the choice of a Skyrme parameter set on the computational results. 
For this purpose, we consider the FNS correction to the ground-state ($1s_{1/2}$) energy and $g$~factor for hydrogen-like lead \ce{^{208}_{82}Pb^{81+}}.
In Table \ref{table:parameters} three different widely used parametrizations (known as LNS, SLy5 and SKP \cite{COLO2013142}) are compared.
In the first three columns of Table \ref{table:parameters} some selected Skyrme parameters are presented in order to illustrate large differences between these parameter sets.
These differences can be seen even more clearly by comparing the values of root-mean-square (RMS) nuclear radius obtained by using each of the parameter sets. 
As a result, the FNS corrections presented in the last two columns also vary significantly in such a way that the results can turn out to be larger or smaller than the FNS corrections obtained in the homogeneously charged sphere approximation (using the RMS radius value of 5.5012 fm for \ce{^{208}_{82}Pb} \cite{ANGELI201369}).

However, it is well known that the magnitude of the FNS correction is highly influenced by the value of RMS nuclear radius \cite{Shabaev_1993, GLAZOV2002408}.
Hence, it is natural to adjust Skyrme parameters to reproduce the experimental RMS radius beforehand and only then calculate the FNS correction.
We found that the RMS radius is most sensitive with respect to varying the Skyrme parameter $t_0$.
In Table \ref{table:adjust} the results of such adjustments in $t_0$ (to obtain $\sqrt{\langle r^2 \rangle}=5.5012 \ \mathrm{fm}$) are shown.

It can be seen that once the value of RMS radius is fixed, the calculated magnitudes of the FNS corrections become stable, despite the significant differences between the parameter sets. 
We tested this observation on a wide range of ions and parametrizations, and we found that the ambiguity in the choice of a Skyrme parameter set was largely suppressed in all cases simply by adjusting the RMS nuclear radius.

All the FNS corrections, presented in the following discussion, were obtained by using the SLy5 parameter set, one of the most widely used parametrizations of the Skyrme force, and the parameter $t_0$ was adjusted to reproduce the experimental values of RMS nuclear radii in each particular case.

\subsection{Energy levels and importance of the RMS radius}\label{sec:FS}

\begin{figure}[!htbp]
\centering
\includegraphics[width=8.55cm]{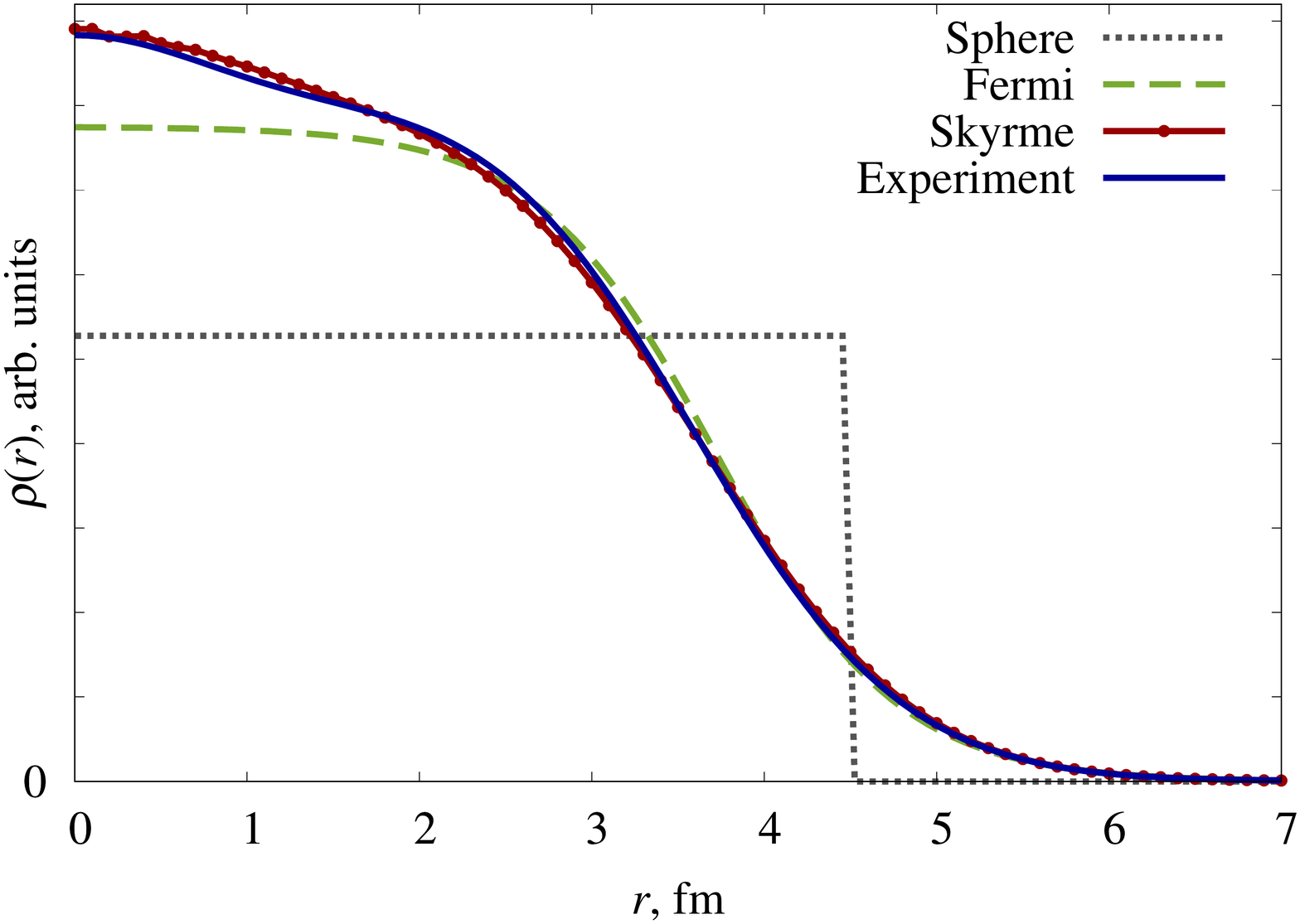}
\caption{(colors online) Comparison between an experimental (``Gauss'') and different model charge distributions for \ce{^{40}_{20}Ca} nucleus. The names of the distributions are explained in Section \ref{sec:density}.}%
\label{fig:Ca_charge}
\end{figure}

\begin{table*}[!htbp]
\caption{Finite nuclear size (FNS) corrections $\Delta E_{\mathrm{FNS}}$ (in units of electron's rest energy) to the energies of the states $1s_{1/2}$, $2s_{1/2}$ and $2p_{1/2}$ for highly charged hydrogen-like ions \ce{^{40}_{20}Ca^{19+}}, \ce{^{116}_{50}Sn^{49+}} and \ce{^{208}_{82}Pb^{81+}}. Different models of the nuclear charge distributions were used in the calculations. The presented calculation uncertainties are due to the experimental uncertainties in RMS nuclear radii \cite{ANGELI201369}. The names of the distributions are explained in Section \ref{sec:density}.}
\label{table:FS}
{\renewcommand{\arraystretch}{1.5}
\renewcommand{\tabcolsep}{0.3cm}
\begin{tabular}{clccc}
\hline \hline
\multirow{2}{*}{\ce{^{40}_{20}Ca^{19+}}} & \multirow{2}{*}{} & $\Delta E_{\mathrm{FNS}}[1s_{1/2}]$ & $\Delta E_{\mathrm{FNS}}[2s_{1/2}]$ & $\Delta E_{\mathrm{FNS}}[2p_{1/2}]$ \\
&& $\times 10^{8}$ & $\times 10^{9}$ & $\times 10^{11}$ \\ 
\hline
& Sphere & 2.8514 & 3.6319 & 1.4696 \\
& Fermi & 2.8502 & 3.6304 & 1.4692 \\
& Skyrme & 2.8502 & 3.6303 & 1.4690 \\
& Bessel & 2.8057 & 3.5737 & 1.4461 \\
& Gauss & 2.8535 & 3.6345 & 1.4708 \\
& Skyrme (+radius unc.) & \phantom{()}2.850(3) & \phantom{()}3.630(4) & \phantom{()}1.469(2) \\
\hline
\multirow{2}{*}{\ce{^{116}_{50}Sn^{49+}}} & \multirow{2}{*}{} & $\Delta E_{\mathrm{FNS}}[1s_{1/2}]$ & $\Delta E_{\mathrm{FNS}}[2s_{1/2}]$ & $\Delta E_{\mathrm{FNS}}[2p_{1/2}]$ \\
&& $\times 10^{6}$ & $\times 10^{7}$ & $\times 10^{8}$ \\
\hline
& Sphere & 3.7906 & 5.3366 & 1.4456 \\
& Fermi & 3.7859 & 5.3299 & 1.4439 \\
& Skyrme & 3.7860 & 5.3301 & 1.4439 \\
& Gauss & 3.7884 & 5.3334 & 1.4448 \\
& Skyrme (+radius unc.) & \phantom{()}3.786(3) & \phantom{()}5.330(4) & \phantom{()}1.444(1) \\
\hline
\multirow{2}{*}{\ce{^{208}_{82}Pb^{81+}}} & \multirow{2}{*}{} & $\Delta E_{\mathrm{FNS}}[1s_{1/2}]$ & $\Delta E_{\mathrm{FNS}}[2s_{1/2}]$ & $\Delta E_{\mathrm{FNS}}[2p_{1/2}]$ \\
&& $\times 10^{4}$ & $\times 10^{5}$ & $\times 10^{6}$ \\
\hline
& Sphere & 1.3172 & 2.2871 & 1.9590 \\
& Fermi & 1.3147 & 2.2827 & 1.9554 \\
& Skyrme & 1.3147 & 2.2827 & 1.9554 \\
& Bessel & 1.3155 & 2.2842 & 1.9566 \\
& Gauss & 1.3155 & 2.2842 & 1.9566 \\
& Skyrme (+radius unc.) & \phantom{(5)}1.3147(4) & \phantom{(5)}2.2827(9) & \phantom{(5)}1.9554(7) \\
\hline \hline
\end{tabular}}
\end{table*}

In Table \ref{table:FS} we present the FNS corrections $\Delta E_{\rm{FNS}}[1s_{1/2}]$, $\Delta E_{\rm{FNS}}[2s_{1/2}]$ and $\Delta E_{\rm{FNS}}[2p_{1/2}]$ calculated by using different nuclear charge distributions for three hydrogen-like ions: \ce{^{40}_{20}Ca^{19+}}, \ce{^{116}_{50}Sn^{49+}} and \ce{^{208}_{82}Pb^{81+}}.
The FNS corrections in the ``Bessel'' and ``Gauss'' rows correspond to experimental charge densities.
Such densities are obtained by expanding $\rho_c(r)$ into sums of spherical zero-order Bessel functions or Gaussians according to Eqs.~\eqref{eq:Bessel}~--~\eqref{eq:Gauss} and fitting the expansion coefficients (as well as any other parameters) to the measured cross sections.
All the values of the fitting parameters used in this paper were taken from Ref.~\cite{DEVRIES1987495}.
We note that for \ce{^{208}_{82}Pb} nucleus two sets of the ``Bessel'' coefficients are known \cite{Fr77b, Eu78}, and for simplicity we present here the results only for the parameters from the Ref. \cite{Eu78}.
The parameters of all the theoretical charge distributions were adjusted to yield the following experimental values of RMS nuclear radii: $\sqrt{\langle r^2 \rangle}=3.4776(19), \ 4.6250(19) \ \rm{and} \  5.5012(13) \ \rm{fm}$ for \ce{^{40}_{20}Ca^{19+}}, \ce{^{116}_{50}Sn^{49+}} and \ce{^{208}_{82}Pb^{81+}}, respectively \cite{ANGELI201369}. 

One peculiar feature can be immediately seen from the results presented in Table \ref{table:FS}: the values obtained in the ``Fermi'' and ``Skyrme'' models agree with each other much better than with the ``experimental values''. 
The explanation for this observation comes from the fact that the value of RMS nuclear radius turns out to be a crucial input parameter, and the experimental charge distributions do not reproduce RMS radii to the current level of precision (as used in the ``Sphere'', ``Fermi'' and ``Skyrme'' calculations). 
This interesting effect can be seen more clearly from the Figure~\ref{fig:Ca_charge}, where we compare different nuclear charge distributions employed in the calculations for \ce{^{40}_{20}Ca^{19+}} ion.
It is instructive to note that despite the fact that the Skyrme and experimental charge distributions are in an excellent agreement with each other, the difference in the corresponding FNS corrections is larger than even between the ``Skyrme'' and ``Sphere'' values.
This surprising result simply comes from the fact that the experimental ``Gauss'' distribution yields $\sqrt{\langle r^2 \rangle}=3.4797 \ \rm{fm}$ instead of 3.4776~fm, and it emphasizes the great influence of the RMS nuclear radius on the magnitude of the FNS effect.
  
The observation described above suggests a straightforward way to estimate the calculation uncertainties for the FNS corrections. Since the RMS nuclear radius turns out to be the main source of uncertainty, one can simply vary the value of the RMS radius within its experimental error bars in the Skyrme model (by varying the $t_0$ parameter) and calculate the corresponding variation in~$\Delta E_{\rm{FNS}}$ or $\Delta g_{\rm{FNS}}$. 
The calculation uncertainties obtained in such a manner are presented in Tables \ref{table:FS} and \ref{table:specific}.

\subsection{$\boldsymbol{g}$ factor and cancellation of the FNS effect in specific differences}\label{sec:specific}

\begin{table*}[!htbp]
\caption{Finite nuclear size (FNS) corrections $\Delta g_{\mathrm{FNS}}$ to the $g$~factors in $1s_{1/2}$, $2s_{1/2}$ and $2p_{1/2}$ states for highly charged hydrogen-like ions \ce{^{40}_{20}Ca^{19+}}, \ce{^{116}_{50}Sn^{49+}} and \ce{^{208}_{82}Pb^{81+}}. In the last two columns the magnitudes of the remaining FNS contribution to the specific differences $g'_{s}$ and $g'_{p}$ are presented. Different models of the nuclear charge distributions were used in the calculations. The presented calculation uncertainties are due to the experimental uncertainties in RMS nuclear radii \cite{ANGELI201369}. The names of the distributions are explained in Section \ref{sec:density}.}
\label{table:specific}
{\renewcommand{\arraystretch}{1.5}
\renewcommand{\tabcolsep}{0.25cm}
\begin{tabular}{clccccc}
\hline \hline
\multirow{2}{*}{\ce{^{40}_{20}Ca^{19+}}} & \multirow{2}{*}{} & $\Delta g_{\mathrm{FNS}}[1s_{1/2}]$ & $\Delta g_{\mathrm{FNS}}[2s_{1/2}]$ & $\Delta g_{\mathrm{FNS}}[2p_{1/2}]$ & $\Delta g'_{\mathrm{FNS}}[1s_{1/2}, 2s_{1/2}]$ & $\Delta g'_{\mathrm{FNS}}[1s_{1/2}, 2p_{1/2}]$ \\
&& $\times 10^{7}$ & $\times 10^{8}$ & $\times 10^{11}$ & $\times 10^{13}$ & $\times 10^{13}$ \\ 
\hline
& Sphere & 1.1316 & 1.4413 & 5.8293 & -2.0 & \phantom{-}0.5 \\
& Fermi & 1.1311 & 1.4407 & 5.7672 & -1.0 & -5.4 \\
& Skyrme & 1.1311 & 1.4406 & 5.8504 & -5.1 & \phantom{-}5.0 \\
& Bessel & 1.1134 & 1.4182 & 5.7560 & \phantom{-}1.5 & \phantom{-}2.5 \\
& Gauss & 1.1324 & 1.4423 & 5.8395 & -2.0 & \phantom{-}1.2 \\
& Skyrme & \multirow{2}{*}{\phantom{()}1.131(1)} & \multirow{2}{*}{\phantom{()}1.441(1)} & \multirow{2}{*}{\phantom{.}5.85(2)} & \multirow{2}{*}{$-$} & \multirow{2}{*}{$-$} \\
& (+radius unc.) &&&&& \\
\hline
\multirow{2}{*}{\ce{^{116}_{50}Sn^{49+}}} & \multirow{2}{*}{} & $\Delta g_{\mathrm{FNS}}[1s_{1/2}]$ & $\Delta g_{\mathrm{FNS}}[2s_{1/2}]$ & $\Delta g_{\mathrm{FNS}}[2p_{1/2}]$ & $\Delta g'_{\mathrm{FNS}}[1s_{1/2}, 2s_{1/2}]$ & $\Delta g'_{\mathrm{FNS}}[1s_{1/2}, 2p_{1/2}]$ \\
&& $\times 10^{5}$ & $\times 10^{6}$ & $\times 10^{8}$ & $\times 10^{10}$ & $\times 10^{10}$ \\
\hline
& Sphere & 1.4426 & 2.0308 & 5.5116 & -7.32 & 3.87 \\
& Fermi & 1.4407 & 2.0282 & 5.5050 & -7.41 & 3.92 \\
& Skyrme & 1.4408 & 2.0282 & 5.5052 & -7.40 & 3.91 \\
& Gauss & 1.4417 & 2.0295 & 5.5086 & -7.41 & 3.92 \\
& Skyrme & \multirow{2}{*}{\phantom{()}1.411(1)} & \multirow{2}{*}{\phantom{()}2.028(2)} & \multirow{2}{*}{\phantom{()}5.505(5)} & \multirow{2}{*}{$-$} & \multirow{2}{*}{$-$} \\
& (+radius unc.) &&&&& \\
\hline
\multirow{2}{*}{\ce{^{208}_{82}Pb^{81+}}} & \multirow{2}{*}{} & $\Delta g_{\mathrm{FNS}}[1s_{1/2}]$ & $\Delta g_{\mathrm{FNS}}[2s_{1/2}]$ & $\Delta g_{\mathrm{FNS}}[2p_{1/2}]$ & $\Delta g'_{\mathrm{FNS}}[1s_{1/2}, 2s_{1/2}]$ & $\Delta g'_{\mathrm{FNS}}[1s_{1/2}, 2p_{1/2}]$ \\
&& $\times 10^{4}$ & $\times 10^{5}$ & $\times 10^{6}$ & $\times 10^{7}$ & $\times 10^{7}$ \\
\hline
& Sphere & 4.5380 & 7.8734 & 6.7814 & -2.271 & 1.138 \\
& Fermi & 4.5292 & 7.8579 & 6.7687 & -2.278 & 1.141 \\
& Skyrme & 4.5291 & 7.8579 & 6.7687 & -2.278 & 1.141 \\
& Bessel & 4.5321 & 7.8630 & 6.7731 & -2.280 & 1.142 \\
& Gauss & 4.5320 & 7.8629 & 6.7730 & -2.280 & 1.142 \\
& Skyrme & \multirow{2}{*}{\phantom{()}4.529(2)} & \multirow{2}{*}{\phantom{()}7.858(3)} & \multirow{2}{*}{\phantom{()}6.769(2)} & \multirow{2}{*}{$-$} & \multirow{2}{*}{$-$} \\
& (+radius unc.) &&&&& \\
\hline \hline
\end{tabular}}
\end{table*}

In general, the same trends as described above for the energy levels hold true also in the case of the FNS corrections to the bound-electron $g$ factor.
In this last section we additionally consider the specific differences of the $g$~factors in $1s_{1/2}$ and $2s_{1/2}$ states, as well as in $1s_{1/2}$ and $2p_{1/2}$ states, for all the charge distributions mentioned above.
These quantities were introduced \cite{spec_dif2002,PhysRevLett.96.253002} with the aim of supressing the FNS effect. Thus, one can expect the specific differences to have more stable values with respect to the choice of nuclear charge distribution.
The specific differences are defined as follows:
\begin{align} 
g'_{s} & = g[\mathrm{2}s_{1/2}] - \xi_{s} g[\mathrm{1}s_{1/2}], \quad  \xi_{s} = \dfrac{\Delta g_{\mathrm{FNS}}[2s_{1/2}]}{\Delta g_{\mathrm{FNS}}[1s_{1/2}]}, \notag \\
g'_{p} & = g[\mathrm{2}p_{1/2}] - \xi_{p} g[\mathrm{1}s_{1/2}], \quad \xi_{p} = \dfrac{\Delta g_{\mathrm{FNS}}[2p_{1/2}]}{\Delta g_{\mathrm{FNS}}[1s_{1/2}]}. 
\end{align}
By expanding the analytical (second-order perturbation theory) expression for $\Delta g_{\mathrm{FNS}}$ from Ref. \cite{GLAZOV2002408} in powers of~$(Z \alpha)$, we obtain:
\begin{align}
\xi_{s} & = \dfrac{1}{8} + 0.110081 (Z \alpha)^{2} + 0.0615871 (Z \alpha)^{4} \notag \\
& + 0.0302009 (Z \alpha)^{6} + 0.0148406 (Z \alpha)^{8} + \{ \mathrm{h.o.} \}, \\
\xi_{p} & = \dfrac{3}{128} (Z \alpha)^{2} + 0.0333355 (Z \alpha)^{4} \notag \\
& + 0.0312421 (Z \alpha)^{6} + 0.0257139 (Z \alpha)^{8} + \{ \mathrm{h.o.} \}.
\end{align}

The calculated values of $\Delta g'_{\mathrm{FNS}} = g'_{\mathrm{ext}} - g'_{\mathrm{point}}$, together with the FNS corrections to the $g$~factors in $1s_{1/2}$, $2s_{1/2}$ and $2p_{1/2}$ states for \ce{^{40}_{20}Ca^{19+}}, \ce{^{116}_{50}Sn^{49+}} and \ce{^{208}_{82}Pb^{81+}}, are shown in Table \ref{table:specific}.
It can be seen that for the specific differences $g'_{s}$ and $g'_{p}$ the FNS effect is indeed suppressed by several orders of magnitude.

However, we also note here that instead of using the analytical expression for $\Delta g_{\mathrm{FNS}}$, one could alternatively evaluate $\xi_{s}$ and $\xi_{p}$ numerically, for example, in the framework of the homogeneously charged sphere approximation.
In this approach, using the new values of $\xi_{s}$ and $\xi_{p}$ for other nuclear models leads to an even stronger suppression of the FNS effect for the specific differences.
For example, in the case of \ce{^{208}_{82}Pb^{81+}} ion, the corrections $\Delta g'_{\mathrm{FNS}}[1s_{1/2}, 2s_{1/2}]$ and $\Delta g'_{\mathrm{FNS}}[1s_{1/2}, 2p_{1/2}]$ within the Skyrme model become only $-1.1 \times 10^{-9}$ and $5.4 \times 10^{-10}$, respectively, which is $2-3$ orders of magnitude smaller than the corresponding values given in Table \ref{table:specific}.
This shows that in the case of heavy ions a direct numerical calculation of $\xi_{s}$ and $\xi_{p}$ (from the best available nuclear model) should be preferred over using analytical formulas.

\section{Conclusions and outlook}\label{sec:conclusion}
We have demonstrated the use of the Skyrme-Hartree-Fock nuclear-structure method as a tool for calculating the finite nuclear size effect in highly charged ions. 
We have shown that, despite the fact that various parametrizations of the Skyrme force differ from each other drastically, the ambiguity in the choice of a parameter set can be significantly suppressed by fixing the value of root-mean-square nuclear radius. 
For this purpose, we suggest adjusting a single Skyrme parameter that has the biggest influence on the value of RMS radius, namely, the parameter $t_0$. 
In this way, the ambiguity associated with the choice of a Skyrme parametrization becomes smaller than the ambiguity stemming from uncertainties in values of nuclear radii.

Our results strongly emphasize the importance of the values of RMS nuclear radii in calculations of FNS corrections.
We have demonstrated that in some cases the value of nuclear radius can be even more important than the shape of the nuclear charge distribution.
In fact, the FNS corrections obtained by means of our approach and by simply using the Fermi distribution agree with each other within the uncertainties in values of nuclear radii.
However, it is clear that the Skyrme model provides a more realistic and thus more reliable description of nuclear charge distributions, which will become crucial in the future when the values of nuclear radii are known to a higher level of accuracy.

\section*{Acknowledgements}
This article comprises parts of the PhD thesis work of Igor Valuev to be submitted to the Heidelberg University, Germany.

\end{document}